# COMBINING THE *SWIFT*/BAT AND THE INTEGRAL/ISGRI OBSERVATIONS

EUGENIO BOTTACINI

*W. W. Hansen Experimental Physics Laboratory and Kavli Institute for Particle Astrophysics and Cosmology, Stanford University, 452 Lomita Mall Stanford, Stanford 94305, California, USA*
*E-mail: eugenio.bottacini@stanford.edu*

MARCO AJELLO

*SLAC National Laboratory and Kavli Institute for Particle Astrophysics and Cosmology, 2575 Sand Hill Road Menlo Park 94025, California, USA*

Current surveys of Active Galactic Nuclei (AGN) find only a very small fraction of AGN contributing to the Cosmic X-ray Background CXB at energies above 15 keV. Roughly 99% of the CXB is so far unresolved. In this work we address the question of the unresolved component of the CXB with the combined surveys of INTEGRAL and *Swift*. These two currently flying X-ray missions perform independent surveys at energies above 15 keV. Our approach is to perform the independent surveys and merge them in order to enhance the exposure time and reduce the systematic uncertainties. We do this with resampling techniques. As a result we obtain a new survey over a wide sky area of 6200 deg$^2$ that is a factor ~4 more sensitive than the survey of *Swift* or INTEGRAL alone. Our sample comprises more than 100 AGN. We use the extragalactic source sample to resolve the CXB by more than a factor 2 compared to current parent surveys.

## 1. Coded mask detectors and current X-ray surveys

Coded-mask instruments start to dominate when X-ray photons are too energetic to be focused with gazing mirrors. Currently flying missions are able to focus X-rays up to 10 keV. At energies above 15 keV the INTEGRAL Soft Gamma-Ray Imager (ISGRI: [1]) and the Burst Alert Telescope (BAT: [2]) coded-mask detectors represent both a major improvement for the imaging of the sky at hard X-ray energies. ISGRI and BAT are flying on board the INTEGRAL [3] and the *Swift* [4] satellites respectively. These two coded-mask detectors are at the forefront of this technology.

The major advantage is that coded-mask instruments have a large field of view and they are therefore very suited for surveys. The ongoing hard X-ray surveys of BAT [5,6] and ISGRI [7,8] continuously shed light onto the properties of the AGN, blazars, Gamma Ray Bursts, Galaxy Clusters and many other astrophysical objects. The drawback of the coded-mask technique is that, by design, ~50% of photons are blocked by the mask and the instrument itself is background dominated, limiting these instruments in sensitivity. This limit becomes evident when determining the contribution of point-like sources to the Cosmic X-ray Background (CXB). The spectrum of the CXB shows a peak ~30 keV [9], where the resolved fraction of point-like sources is a mere ~1%. While at lower energies (~1 keV) focusing X-ray telescopes (e.g. Chandra, XMM-Newton) resolve 100% of the CXB. As a consequence, the sources contributing to the peak of the CXB are undetected.

## 2. Combining the ISGRI and the BAT observations

In [7] the authors compare the exposure of the BAT detected sources in the ISGRI instrument that are detected and un-detected by ISGRI. They conclude that the none-detection by ISGRI is just due to the low exposure of the sources in the instrument. The lower exposure of these sources in the ISGRI instrument is mainly due to the different pointing strategies of the two satellites. BAT is quasi-randomly pointing the sky (and thus covering uniformly the whole sky) while INTEGRAL performs predetermined pointing spending a considerable amount of time observing the Galactic center.

A natural way of increasing the exposure on the hard X-ray sky is obtained by combining the BAT and the ISGRI observations. Indeed, it is possible to merge the independently produced sky survey images by BAT and ISGRI to obtain a much deeper X-ray survey image. We first perform the two independent surveys of BAT and ISGRI in the energy range $18 - 55$ keV. The BAT survey is performed following the recipe as in [5]. ISGRI data are processed using the official INTEGRAL Off-line Science Analysis (OSA) software [10] accounting for bright sources and their variability and time-dependent background models. The mosaic images are resampled using a bilinear interpolation obtaining new aligned images of the same pixel size. The newly obtained images are cross-calibrated and finally merged. The advantage of this method is not only to reduce the statistical noise due to increase exposure time, but also the heavy systematic errors that affect coded-mask detectors [11] are reduced being the systematic uncertainties of the two instruments not correlated.

We have applied this technique to an extragalactic sky area of 6200 deg$^2$. To study the quality of the final mosaic image we investigate the pixel significance distribution (see Figure 1) that is the ratio of the intensities and their associated errors of the mosaic image. The significance histogram follows a Gaussian distribution with zero mean and unitary variance. The errors of the final mosaic are computed propagating the errors of both BAT and ISGRI surveys considering that the covariance term between the both surveys is zero being their associated errors uncorrelated. This survey is complete to a flux of $\sim 1.2 \times 10^{-11}$ erg cm$^{-2}$ s$^{-1}$ and reaches a flux limit down to $\sim 3 \times 10^{-12}$ erg cm$^{-2}$ s$^{-1}$. Setting the source detection threshold to 4.8 sigma we detect more than 100 AGN, galaxy clusters, galaxies, and Galactic sources.

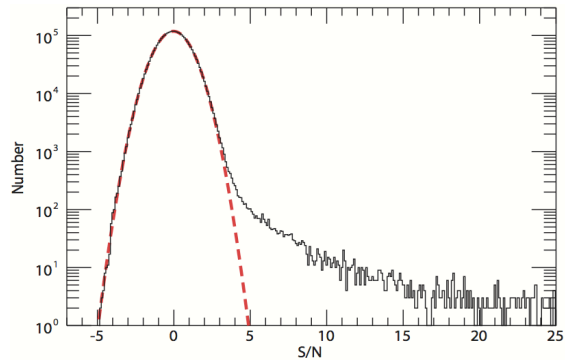

Figure 1. The distribution of the pixel significances of the merged mosaic image of 6200 deg$^2$. The red dashed line is a fitted Gaussian distribution with dispersion $\sigma = 1.0$. There are no wings in the distribution. The noise is under control. Positive significances above the value 4.8 represent real detected sources.

## 3. Conclusions

We have combined the BAT and ISGRI surveys in the energy range $18 - 55$ keV. This allows one to greatly enhance the exposure on the hard X-ray sky. In turn statistical and systematic uncertainties are well behaved. This allows sampling fluxes to a limit of $\sim 3 \times 10^{-12}$ erg cm$^{-2}$ s$^{-1}$. As a result we detect more than 100 AGN over a sky area of 6200 deg$^2$. This permits resolving the CXB by more than a factor 2 compared to the sample in [12].

**Acknowledgments**

We thank the INTEGRAL and the *Swift* team for the observations and the support. E.B. acknowledges support through SAO grant GO1-12144X.